\documentclass[english]{article}
\usepackage[T1]{fontenc}
\usepackage[latin9]{inputenc}
\usepackage{geometry}
\usepackage{amsmath}
\usepackage{graphicx}
\usepackage{amssymb}
\usepackage{esint}

\makeatletter

\providecommand{\tabularnewline}{\\}

\newcommand{\lyxaddress}[1]{
\par {\raggedright #1
\vspace{1.4em}
\noindent\par}
}

\makeatother

\usepackage{babel}

\begin{document}

\title{Adaptive propagation of quantum few-body systems with time-dependent
Hamiltonians}

\author{J. C. Cremon}

\maketitle

\lyxaddress{Mathematical Physics, LTH, Lund University, SE-22100 Lund, Sweden}
\begin{abstract}
In this study, a variety of methods are tested and compared for the
numerical solution of the Schrödinger equation for few-body systems
with explicitely time-dependent Hamiltonians, with the aim to find
the optimal one. The configuration interaction method, generally applied
to find stationary eigenstates accurately and without approximations
to the wavefunction's structure, may also be used for the time-evolution,
which results in a large linear system of ordinary differential equations.
The large basis sizes typically present when the configuration interaction
method is used calls for efficient methods for the time evolution.
Apart from efficiency, adaptivity (in the time domain) is the other
main focus in this study, such that the time step is adjusted automatically
given some requested accuracy. A method is suggested here, based on
an exponential integrator approach, combined with different ways to
implement the adaptivity, which was found to be faster than a broad
variety of other methods that were also considered.
\end{abstract}

\section{\label{sec:Introduction}Introduction}

Experimental development in recent years has allowed for creation
and detailed control of tunable quantum mechanical few-body systems.
For example, trapped ultra-cold atomic or molecular gases are systems
where both trapping potentials and the particle-particle interactions
may be directly controlled via externally applied fields \cite{wieman-pritchard-wineland-1999-RMP,dalfovo-giorgini-pitaevskii-stringari-1999-RMP,bloch-2005-NatPhys,bloch-dalibard-zwerger-2008-RMP,giorgini-pitaevskii-stringari-2008-RMP},
and such systems have also been realized with very few particles \cite{bloch-dalibard-zwerger-2008-RMP,jochim-2011}.
Another example is electrons confined in so-called quantum dots, small
semiconductor based structures, where for example electric and magnetic
fields also can be used to affect the particles \cite{reimann-manninen-2002-RMP,samuelson-et-al-2004-PhysicaE,hanson-et-al-2007-RMP}.
Such systems can allow for the study of, for example, non-equilibrium
time-dependent processes, quantum quenches, or laser-induced dynamics,
with strongly interacting particles.

In this study, a variety of numerical methods have been tested, to
simulate the time-evolution of such systems as efficiently as possible.
A method based on an exponential-integrator approach was found to
be the most efficient. It re-uses some important concepts from one
presented in a work by Park and Light \cite{park-light-1986}, but
is here extended to handle also time-dependent Hamiltonians, by implementing
the time-adaptivity in a different way.

The methods are described in section \ref{sec:Methods}. Then a physical
test case, along with the numerical performance tests, is presented
in section \ref{sec:Results}. A summary is given in section \ref{sec:Summary},
and the appendix contains some remarks on the implementation, along
with a short discussion about some other methods.

\section{\label{sec:Methods}Method}

The configuration interaction method (or {}``exact diagonalization'')
is widely used to model quantum mechanical few-body systems. Its versatility
allows its usage in different fields such as quantum chemistry, nuclear
physics, and condensed matter physics (see for example Refs. \cite{cramer-2004-book,navratil-quaglioni-stetcu-barrett-2009-JPG,rontani-et-al-2006}).
In practice the method is limited to very small numbers of particles,
but its advantages are that it does not involve any initial assumptions
about the structure of the many-body wavefunction, and it is fully
capable to describe also excited states. Furthermore, it is rather
flexible in the sense that it can handle both bosons and fermions,
and is not restricted to any particular kind of interaction between
the particles. While typically employed to find eigenstates for the
time-independent Schrödinger equation, it can also be used as a foundation
for solving the time-dependent one,

\begin{equation}
\frac{d}{dt}\vert\Psi(t)\rangle=-iH(t)\vert\Psi(t)\rangle\label{eq:td-schrodinger}\end{equation}
with some initial state $\vert\Psi(0)\rangle$. (Throughout this paper
we set $\hbar=1$.) The states are expanded as linear combinations
of some set of (static) many-body basis states, denoted by $\vert\Phi_{\nu}\rangle$
below. Eigenstates of a Hamiltonian may be obtained by finding eigenvectors
of its matrix representation in the given basis. For many typical
cases this matrix is very sparse. For a state $\vert\Psi(t)\rangle$
that changes in time, the time-dependence is contained in the coefficients
$b_{\nu}(t)$ of the expansion:

\begin{equation}
\vert\Psi(t)\rangle=\sum_{\nu}b_{\nu}(t)\vert\Phi_{\nu}\rangle\label{eq:linear-expansion}\end{equation}
With this expansion inserted in Eq.~(\ref{eq:td-schrodinger}), it
yields a system of linearly coupled ordinary differential equations
for the coefficients $b_{\nu}(t)$. There is then a variety of numerical
approaches that can be used to solve it, although the large dimensionality
of typical problems makes it crucial to work with efficient methods.
Within this study several options have been compared, in an attempt
to find the optimal one, for the systems intended to be investigated.
For example, standard Runge-Kutta methods, explicit or implicit, may
be used. However, another class of methods, generally referred to
as exponential integrators (described in the following sections),
have in many cases been found to be more efficient for the problem
at hand \cite{hochbruck-ostermann-2010}.

A number of earlier studies have utilized time-dependent configuration
interaction methods, in particular to study electronic response to
applied laser fields (see, for example, Refs.~\cite{tdci-klamroth-2003,tdci-larsen-schwartz-2003,tdci-krause-klamroth-saalfrank-2005,tdci-rohringer-gordon-santra-2006,tdci-krause-klamroth-saalfrank-2007,tdci-pabst-greenman-ho-mazziotti-santra-2011,tdci-hochstuhl-bonitz-2012}).
Typically in this context, a Hartree-Fock solution together with single
particle-hole excitations is used to model the excitation of an atom
or molecule by a light field. Some other recent studies suggest and
use time-dependent methods specialized for interacting bosons, such
as a multiconfigurational Hartree method \cite{alon-streltsov-cederbaum-2008},
or, in another study, a method employing time-dependent basis states,
which results in a set of nonlinear equations \cite{streltsov-alon-cederbaum-2007}.
In the present study, however, the wish is to handle systems with
as general time-dependent Hamiltonians as possible, including time-dependent
interactions, and the possibility for both bosons and fermions.

The methods tested here are described in the sections below: Section
\ref{sub:Methods-const} describes propagation of constant (time-independent)
Hamiltonians, closely following the approach by Park and Light \cite{park-light-1986}.
But for time-dependent Hamiltonians, a different method (in two variants)
is suggested and tested in this study, described in section \ref{sub:Methods-TD}.
A high-order Runge-Kutta method is also considered here, described
in section \ref{sub:Methods-RK}. Initially in this project, an implicit
method was also tested but found to be inefficient, details about
this are given in section \ref{sub:Implicit-methods} of the appendix.

\subsection{\label{sub:Methods-const}Adaptive Lanczos propagation with a constant
Hamiltonian (ALC)}

If the Hamiltonian $H$ in Eq.~(\ref{eq:td-schrodinger}) does not
depend on time, then a formal solution is given by

\begin{equation}
\vert\Psi(t)\rangle=\exp(-itH)\vert\Psi(0)\rangle\label{eq:notd-exp-solution}\end{equation}
with a so-called exponential integrator. Because of the large dimensions
of the Hilbert spaces considered here, it is not feasible to directly
compute the whole matrix exponential. Instead, the Lanczos process
\cite{lanczos-1950,golub-vanloan} can be used to create an approximation
for $H$, which may then be used to compute the evolution during a
small time step $\Delta t$, and this scheme is then iterated several
times. This approach was presented in a work by Park and Light \cite{park-light-1986}.
The method, with a small modification, is described in detail below
since its main concepts are relevant also for time-dependent Hamiltonians,
as discussed in the next section. For a more mathematical perspective
on exponential integrators, see for example Refs.~\cite{saad-1992,hochbruck-lubich-1997,hochbruck-lubich-selhofer-1998,hochbruck-lubich-BIT-1999}.

It should be noted here that even if the method below is only applicable
for Hamiltonians without an explicit time-dependence, it can of course
be readily applied to any piece-wise constant Hamiltonian by separating
the time axis into appropriate subintervals.

The Lanczos process, truncated after some finite number of iterations,
yields an orthogonal set of vectors in a matrix $K$, the Krylov space,
coupled by a tridiagonal matrix $T$ \cite{lanczos-1950,golub-vanloan}:

\begin{equation}
H=KTK^{\dagger}=\left[\begin{array}{cccc}
\vert & \vert & \vert\\
\vert k_{0}\rangle & \vert k_{1}\rangle & \vert k_{2}\rangle & \cdots\\
\vert & \vert & \vert\end{array}\right]\left[\begin{array}{cccc}
\alpha_{0} & \beta_{0} & 0 & 0\\
\beta_{0} & \alpha_{1} & \beta_{1} & 0\\
0 & \beta_{1} & \alpha_{2} & \ddots\\
0 & 0 & \ddots & \ddots\end{array}\right]\left[\begin{array}{ccc}
- & \vert k_{0}\rangle^{\dagger} & -\\
- & \vert k_{1}\rangle^{\dagger} & -\\
- & \vert k_{2}\rangle^{\dagger} & -\\
 & \vdots\end{array}\right]\label{eq:lanczos-factorization}\end{equation}
The first vector $\vert k_{0}\rangle$ is just $\vert\Psi(t_{0})\rangle$,
and $t_{0}$ is the time at the beginning of the small time step.
Then, the next vector $\vert k_{1}\rangle$ is obtained by applying
$H$ on $\vert k_{0}\rangle$ and projecting out the component along
$\vert k_{0}\rangle$, to make $\vert k_{1}\rangle$ orthogonal to
$\vert k_{0}\rangle$, and similarly for subsequent vectors. Only
matrix-vector products are required, no other manipulation of $H$
is needed, making this approach useful for sparse matrices. By inserting
the factorization above into Eq.~(\ref{eq:notd-exp-solution}), together
with a diagonalization $T=QDQ^{\dagger}$, we get \begin{equation}
\vert\Psi(t_{0}+\Delta t)\rangle=KQe^{-i\Delta tD}Q^{\dagger}K^{\dagger}\vert\Psi(t_{0})\rangle\label{eq:notd-exp-solution-lanczos}\end{equation}
As $D$ is diagonal, it can be exponentiated trivially, and $\vert\Psi(t_{0}+\Delta t)\rangle$
can be computed.

It remains however to choose the dimension $d_{K}$ of the Krylov
space, and the time step $\Delta t$. The approach here is very similar
to the one described in Ref. \cite{park-light-1986} for $\Delta t$.
With a closer look at Eq.~(\ref{eq:notd-exp-solution-lanczos}),
one may view the expression as a differential equation {}``within
the Krylov space''. That is, for a given dimension $d_{K}$, the
state $\vert\Psi(t)\rangle$ is allowed to evolve only within the
space -- which has a very small dimensionality compared to the full
system. More explicitly, one has the tridiagonal system \begin{equation}
\frac{d}{dt}\bar{c}=-iT\bar{c}\label{eq:tridiagonal-diff-eq}\end{equation}
where $c_{j}(t)=\langle k_{j}\vert\Psi(t)\rangle$. Initially, we
have $c_{0}(t_{0})=1$ and all other components are zero, since $\vert k_{0}\rangle=\vert\Psi(t_{0})\rangle$,
and all Krylov vectors are orthogonal to each other. The tridiagonal
structure of $T$ implies that, as time evolves, $c_{1}(t)$ should
grow large before $c_{2}(t)$ does. The last component, denoted $c_{\mathrm{last}}(t)$,
should only grow significantly large after some time has passed. On
the other hand, if it does become large it implies that also the subsequent
components could be significant -- if they had actually been calculated.
Based on this, $\Delta t$ can be adjusted adaptively at each time
step to keep $c_{\mathrm{last}}(t_{0}+\Delta t)$ smaller than some
fixed tolerance $\epsilon$, but as close to this value as possible.
An estimate for $\Delta t$ based on perturbation theory is given
in the work by Park and Light \cite{park-light-1986}. But in the
present implementation it was instead determined using recursive bisection
of the interval $[t_{0},t_{\mathrm{end}}]$, where $t_{\mathrm{end}}$
is the user-specified end time of the simulation, to make sure the
chosen $\Delta t$ is really optimal.

The dimension $d_{K}$ of the Krylov space is set to some fixed value
throughout the time evolution. Generally, a larger space should give
higher accuracy as more basis vectors are available, and thereby allow
for larger (and fewer) time steps to be taken. But more memory is
required to store the vectors, and the inherent numerical instability
of the Lanczos process \cite{golub-vanloan} might affect the result
for large $d_{K}$. Earlier studies and implementations of exponential
integrators have found a dimension in the range 20--30 to be most
efficient \cite{park-light-1986,sidje-1998,edwards-et-al-1994,bergamaschi-et-al-2006}.
After some testing, that choice was found to be good here as well,
and $d_{K}=30$ was used to produce the results reported below. A
smaller value ($d_{K}\sim10$) gave considerably worse performance,
while a larger value ($d_{K}\sim100$) did not improve it much.

\subsection{\label{sub:Methods-TD}Adaptive Lanczos propagation for time-dependent
Hamiltonians}

Exponential integrators for explicitly time-dependent matrices are
also covered in the literature, see for example Refs. \cite{hochbruck-lubich-BIT-1999,magnus-1954,wilcox-1967,fer-1958,klarsfeld-oteo-1989,hochbruck-lubich-2003-magnus,kosloff-1994}.
However, not many studies are available that deal with adaptive step-size
control for the case when the matrix is very large and sparse, such
that the Lanczos process needs to be used.

The method described in the previous section is unfortunately not
applicable here, since Eq. (\ref{eq:notd-exp-solution}) is not valid
if the Hamiltonian is explicitely time-dependent. For small time steps
$\Delta t$, the following expression can be used as an approximation:

\begin{equation}
\vert\Psi(t_{0}+\Delta t)\rangle=\exp\biggl(\Omega_{1}(t_{0},\Delta t)\biggl)\vert\Psi(t_{0})\rangle\label{eq:td-exp-solution-approx-magnus-1st}\end{equation}
with \begin{equation}
\Omega_{1}(t_{0},\Delta t)=-i\intop_{t_{0}}^{t_{0}+\Delta t}H(\tau)d\tau\label{eq:Omega1}\end{equation}
One may then use $\Omega_{1}$ to construct a Krylov space and compute
the time evolution for a small $\Delta t$, similar to the approach
described above for a time-independent Hamiltonian. But the previous
approach to afterwards tune $\Delta t$ given a certain Krylov space
of some fixed dimension is no longer possible. The problem is that
the matrix used to build the Krylov space now itself depends on $\Delta t$.
And even if one could do this, it is not clear how large an error
is made in the effective averaging of $H(t)$ during the time interval.
A solution is to compute two different approximations to $\vert\Psi(t_{0}+\Delta t)\rangle$,
where one is known to be more accurate than the other. Then, if the
error between them is small enough, one can assume that the approximation
is good -- otherwise, the time step $\Delta t$ must be decreased.
(The error is here defined as the Euclidian norm of the difference
of the vectors.)

Below, two different schemes (denoted AL1 and AL2) are presented which
both provide suitable pairs of approximations, followed by a discussion
of how the time step $\Delta t$ and the Krylov dimension are finally
adjusted in the implementation. Some details about the actual integration
of the Hamiltonian, as needed in the computation of $\Omega_{1}$,
are given in section \ref{sub:Implementation-details} of the appendix.

\subsubsection{\label{sub:Methods-TD-AL1}Step-doubling (AL1)}

One way to get two different approximations of $\vert\Psi(t_{0}+\Delta t)\rangle$
is to first compute an approximation as mentioned above, and then
compute another one by instead taking two steps of length $\frac{\Delta t}{2}$.
The latter should be more accurate since a smaller time step is used,
and this solution is also used to continue the simulation. This approach
is commonly referred to as {}``step-doubling'', see e.g. the discussion
in Ref. \cite{numerical-recipes} about adaptive Runge-Kutta methods.

\subsubsection{\label{sub:Methods-TD-AL2}The Magnus expansion (AL2)}

Another way is to use the Magnus expansion \cite{magnus-1954}, in
which a series of correctional terms is added to the exponent in Eq.~(\ref{eq:td-exp-solution-approx-magnus-1st}).
The second order Magnus expansion states a more accurate approximation
as\begin{equation}
\vert\Psi(t_{0}+\Delta t)\rangle=\exp\biggl(\Omega_{1}(t_{0},\Delta t)+\Omega_{2}(t_{0},\Delta t)\biggl)\vert\Psi(t_{0})\rangle\label{eq:td-exp-solution-approx-magnus-2nd}\end{equation}
with

\begin{equation}
\Omega_{2}(t_{0},\Delta t)=-\frac{1}{2}\intop_{t_{0}}^{t_{0}+\Delta t}d\tau_{1}\intop_{t_{0}}^{\tau_{1}}d\tau_{2}[H(\tau_{1}),H(\tau_{2})]\label{eq:Omega2}\end{equation}
where the commutator $[A,B]=AB-BA$ is used. Higher order terms of
the Magnus expansion involve increasingly nested commutators.

In the same spirit as the Magnus expansion, the Fer expansion \cite{fer-1958}
and the Wilcox approach \cite{wilcox-1967} also provide refined approximations,
though the Magnus solution appears to be more widely used \cite{hochbruck-ostermann-2010,klarsfeld-oteo-1989}.
Their first order versions all correspond to Eq.~(\ref{eq:td-exp-solution-approx-magnus-1st})
\cite{klarsfeld-oteo-1989}. Some successful uses of the Magnus expansion
together with the Lanczos process have been made previously. For example,
in Ref. \cite{tannor-besprozvannaya-williams-1992} a way to evaluate
the commutator within the Krylov space is considered, when the calculations
are done in the so-called interaction picture, where the Hamiltonian
is taken to be separable in a dominant part which is, in some sense,
easier to handle, and a {}``smaller'' but more difficult part. Another,
more recent, study is the one in Ref.\emph{ }\cite{kormann-et-al-2008},
where for some particular special cases a commutator-free method is
derived. In any case, in this study the aim was to find a robust way
to handle fairly general Hamiltonians.

The matrix $(\Omega_{1}+\Omega_{2})$ can be used to build a Krylov
space and perform the time evolution, as described above. Then, a
solution obtained using only $\Omega_{1}$ should be less accurate,
and the difference between them provides an estimate of the local
error.

\subsubsection{\label{sub:Methods-TD-adjusting-dt}Adjusting the time step $\Delta t$}

With a pair of approximations $\vert\Psi_{1}(t_{0}+\Delta t)\rangle$
and $\vert\Psi_{2}(t_{0}+\Delta t)\rangle$, obtained by one of the
schemes described above, where it is clear that one of the approximations
is better than the other, the local error can be estimated as $\Vert\vert\Psi_{1}\rangle-\vert\Psi_{2}\rangle\Vert_{2}$.
If this error is less than some chosen tolerance $\epsilon$, the
solution known to be more accurate is accepted and the simulation
can continue. Otherwise, the whole time step is re-done using the
step size $\Delta t/2$.

Assuming that a time step is successful, with a sufficiently small
error, one can then try to estimate what would be the optimal $\Delta t$
for the following time step. Many implementations of adaptive Runge-Kutta
methods exploit detailed knowledge about the convergence properties
of the method to make sophisticated estimates of the optimal length
of the next time step \cite{numerical-recipes,dormand-prince-5-4,dormand-prince-8-7}.
Here a simpler approach is taken: If the error is less than $\frac{\epsilon}{2}$
we increase the next time step length by a factor of $1.1$, since
we apparently were using an unnecessary small $\Delta t$ in the last
step, for the given tolerance. And if the error is larger than $\frac{\epsilon}{2}$
we instead divide the step length by $1.1$, to reduce the risk that
the error becomes to large in the next step, which would then have
to be re-done.

\subsubsection{\label{sub:Methods-TD-adjusting-dK}Adjusting the Krylov dimension
$d_{K}$}

The dimension of the Krylov space is the other important parameter.
Since the exponent used to build the Krylov space depends on $\Delta t$,
it is not possible to adjust the step length afterwards in order to
optimally use the obtained space, as was done for constant Hamiltonians
in section \ref{sub:Methods-const}. However, with $\Delta t$ fixed
only a finite number of Krylov vectors will actually be needed during
the time step; that is, the coefficients $c_{j}$ in Eq.~(\ref{eq:tridiagonal-diff-eq})
should be very small for sufficiently large indices $j$. Thus, for
every new Krylov vector that is generated, the resulting coefficient
$c_{\mathrm{last}}(t_{0}+\Delta t)$ is computed. If its magnitude
is larger than some tolerance, more Krylov vectors are needed. But
if it is small enough the obtained Krylov space should be sufficient
for this time step. In this way, the dimension of the Krylov space
is not fixed but always adjusted to suit the current step size.

However, practical memory constraints, and the possible numerical
instability in the Lanczos process, suggests that the Krylov space
should not be allowed to grow too large. Thus a parameter $d_{K,\max}$
is used: If the Krylov space reaches this dimension, without the coefficient
$c_{\mathrm{last}}(t_{0}+\Delta t)$ being small enough, the whole
time step is re-done with step size $\frac{\Delta t}{2}$. Furthermore,
in an attempt to avoid the maximum dimension being reached, the next
step length is divided by $1.1$ if the previously needed Krylov dimension
was larger than $0.8\cdot d_{K,\max}$. Similarly to the case with
a constant Hamiltonian (see section \ref{sub:Methods-const}), $d_{K,\max}=30$
was found to be a good choice.

\subsection{\label{sub:Methods-RK}An adaptive Runge-Kutta method (RK8)}

In order to check the performance of the exponential integrator methods,
an adaptive explicit Runge-Kutta method was also implemented. A method
presented by Prince and Dormand \cite{dormand-prince-8-7} (denoted
by RK8 in the present article) was found to be the most efficient
for the test case in this study (see below). This method uses 13 matrix-vector
multiplications per time step, and produces a solution accurate to
8:th order in the step size, along with an embedded 7:th order approximation
that can be used to estimate the local error, which is then required
to be smaller than some value $\epsilon$. The present implementation
follows very closely the recipe in Ref.~\cite{dormand-prince-8-7},
with the only exception that the error is, in the present work, estimated
using the Euclidian norm ($\Vert\cdot\Vert_{2}$) instead of the maximum
norm ($\Vert\cdot\Vert_{\infty}$).

\section{\label{sec:Results}Results and discussion}

The physical test case is first presented in section \ref{sub:Results-TestCase},
along with a short discussion about the resulting dynamics. Then,
the numerical test results are given in section \ref{sub:Results-TestResults}.

\subsection{\label{sub:Results-TestCase}The test case: Interacting bosons in
a 1D confinement}

To demonstrate the methods discussed above, a test system was chosen,
intended to be a few-body system with fairly strong correlations between
the particles. A few bosonic particles trapped in an one-dimensional
quantum well with infinitely high walls were considered. They interact
via a repulsive short-range delta-interaction. This could model ultra-cold
atoms trapped in a quasi-one-dimensional waveguide, interacting via
van der Waals interactions. Such systems have been realized in several
experiments \cite{bloch-dalibard-zwerger-2008-RMP}. With the particle
mass, the width of the well, and Planck's constant $\hbar$ all set
to one, the Hamiltonian is

\begin{equation}
H(t)=\sum_{j=1}^{N}\biggl(-\frac{1}{2}\frac{d^{2}}{dx_{j}^{2}}+f(t)x_{j}\biggl)+\frac{1}{2}\sum_{j\neq k}^{N}g\delta(x_{j}-x_{k})\label{eq:test-hamiltonian}\end{equation}
where the particles are then confined to the interval $0<x<1$. The
interaction strength is set to $g=2$, which gives a fairly strong
repulsive interaction, such that the bosons should not be expected
to form a Bose-Einstein condensate, but neither avoid each other completely
(which would correspond to a so-called Tonks-Girardeau gas \cite{girardeau-1960}).

In order to introduce some dynamics in the system, the otherwise flat
bottom of the well is tilted, with the addition of the potential energy
term $f(t)x$, where the function $f(t)$ contains all explicit time
dependence of the Hamiltonian. Initially, for times $t<0$, the system
is {}``prepared'' in the ground state for $f(t)=100$, such that
the particle cloud is essentially located in the left corner of the
well -- this is the initial state $\vert\Psi(0)\rangle$. Then, the
system is evoluted in time in the interval $0\le t\le10$, with $f(t)$
varying as described in section \ref{sub:Results-TestCase-TDPotentials}
below. To illustrate the dynamics of the system, the average particle
position is plotted in Fig.~\ref{fig:results-oscillations} as a
function of time; defined as $x_{\mathrm{mean}}(t)=\langle\Psi(t)\vert\frac{1}{N}\sum_{j=1}^{N}x_{j}\vert\Psi(t)\rangle$.

In the results below, $N=5$ particles were put into the system. From
the physics point of view, it is interesting to see the effect of
the particle-particle interaction, and for this reason the simulation
is also done with the interaction turned off (that is, with $g=0$).

\subsubsection{\label{sub:Results-TestCase-TDPotentials}Time-dependent potentials}

Four different time-dependent potentials are considered in the tests
here; or, rather, four different choices of the function $f(t)$ discussed
above. The intention is to test how the methods handle different situations.
They are described below, and denoted \emph{a}, \emph{b}, \emph{c}
and \emph{d}. The different functions are also plotted in Fig.~\ref{fig:results-oscillations}.

\emph{a}) To test a system without an explicit time-dependence in
the Hamiltonian, at times $t>0$ we let \[
f_{a}(t)=0\]
As the system is initially in the ground-state for $f_{t<0}(t)=100$,
this sudden release of the particles will create oscillations in the
system.

\emph{b}) In this case a sawtooth function was considered, defined
by\[
f_{b}(t)=\begin{cases}
0\le t<5 & 100\cdot(1-0.2t)\\
5\le t\le10 & 100\cdot(1-0.2[t-5])\end{cases}\]

\emph{c}) Many physically interesting situations involve oscillating
potentials, for example electrons coupled to an electromagnetic field
which may excite the system. Since the physical response may be very
frequency dependent, a function was chosen to cover a broad range
of frequencies:\[
f_{c}(t)=\begin{cases}
0\le t<5 & 10\cdot\cos(2\pi t^{2})\\
5\le t\le10 & 10\cdot\cos(2\pi[t-5]^{2})\end{cases}\]
In this way the angular frequency can be said to increase from zero
up to $2\pi\cdot5\approx30$, which covers the energy difference between
the ground state and first excited state of the infinite well potential,
which is $3\pi^{2}/2\approx15$. For this particular test case, the
initial state $\vert\Psi(0)\rangle$ was computed for the potential
with $f_{t<0}(t)=10$.

\emph{d}) To test with a more strongly oscillating potential, $f_{d}(t)$
was chosen identical to $f_{c}(t)$ defined above, except that the
prefactors were changed to 100 instead of 10.

\subsubsection{\label{sub:Results-TestCase-BasisStates}Basis states}

As a basis, the single-particle orbitals of the infinite well are
used, i.e. the functions $\phi_{n}(x)=\sqrt{2}\sin(\pi nx)$ with
$n\geq1$. These orbitals are then used to build a space of many-particle
states (the Hilbert space). Because of computational limitations the
basis must be truncated, here this is done by only including a finite
number $d_{1}$ of single-particle orbitals ($10$, $20$ or $30$
are used here with all methods). A larger number of orbitals give
higher spatial accuracy. As shown by the basis sizes presented in
Table~\ref{tab:results-data} below (the column marked with $d_{N}$)
the problem dimension grows rapidly with both the number of particles
and the number of orbitals. Also, $d_{N}$ grows very quickly as a
function of $N$.

\subsubsection{\label{sub:Results-TestCase-ResultingDynamics}Resulting dynamics}

The four different test cases result in qualitatively different dynamics;
the result is shown in Fig.~\ref{fig:results-oscillations}. To be
able to distinguish which features that originate from the motion
of single particles, and which that are interaction effects, the simulation
is done for both $g=0$ and $g=2$ (where $g$ is the interaction
strength). A general trend that can be seen in all four cases is that
the repulsive interaction has a limiting effect on the amplitude of
the oscillations.

\emph{a}) For this case, where the tilted potential is suddenly released
at $t=0$, oscillations are seen in the system. Essentially, the particles
are bouncing back and forth against the edges of the confinement.
The interaction decreases the amplitude of the oscillations, but it
also appears to introduce some new oscillation with another frequency,
with the overall effect of a beating mode.

\emph{b}) When the tilted potential is modulated with a sawtooth function,
the dynamics are initially very different. Now, the particles are
not suddenly released but slowly let to expand, and during the interval
$0<t<5$ there is practically no oscillation to be seen -- a so-called
adiabatic evolution where the system is always in the instantaneous
ground state of the Hamiltonian \cite{born-fock-1928}. Then, at time
$t=5$ the sawtooth function {}``kicks'' the system and introduces
oscillations, which after a transient period resemble those seen for
the previous case (\emph{a}).

\emph{c}) In this case an oscillating potential is considered, which
can be expected to excite the system as discussed in section \ref{sub:Results-TestCase-TDPotentials}.
Although this induces a lot of noise in the particles' oscillations,
in general the dynamics show similarities with case \emph{a}.

\emph{d}) The oscillating potential is here much stronger than in
the previous case (\emph{c}), but otherwise the same. It appears that
the potential is so strong that it completely drives the system --
at least during the interval $0<t<5$ the particles seem to directly
follow the changes in the potential. Then, at time $t=5$ the dynamics
are more difficult to interpret. The single-particle oscillations
are quite fast, and also show a beating pattern, while these oscillations
are again damped by the repulsive interaction.

\begin{center}
\begin{figure}
\noindent \begin{centering}
\includegraphics[width=0.5\textwidth]{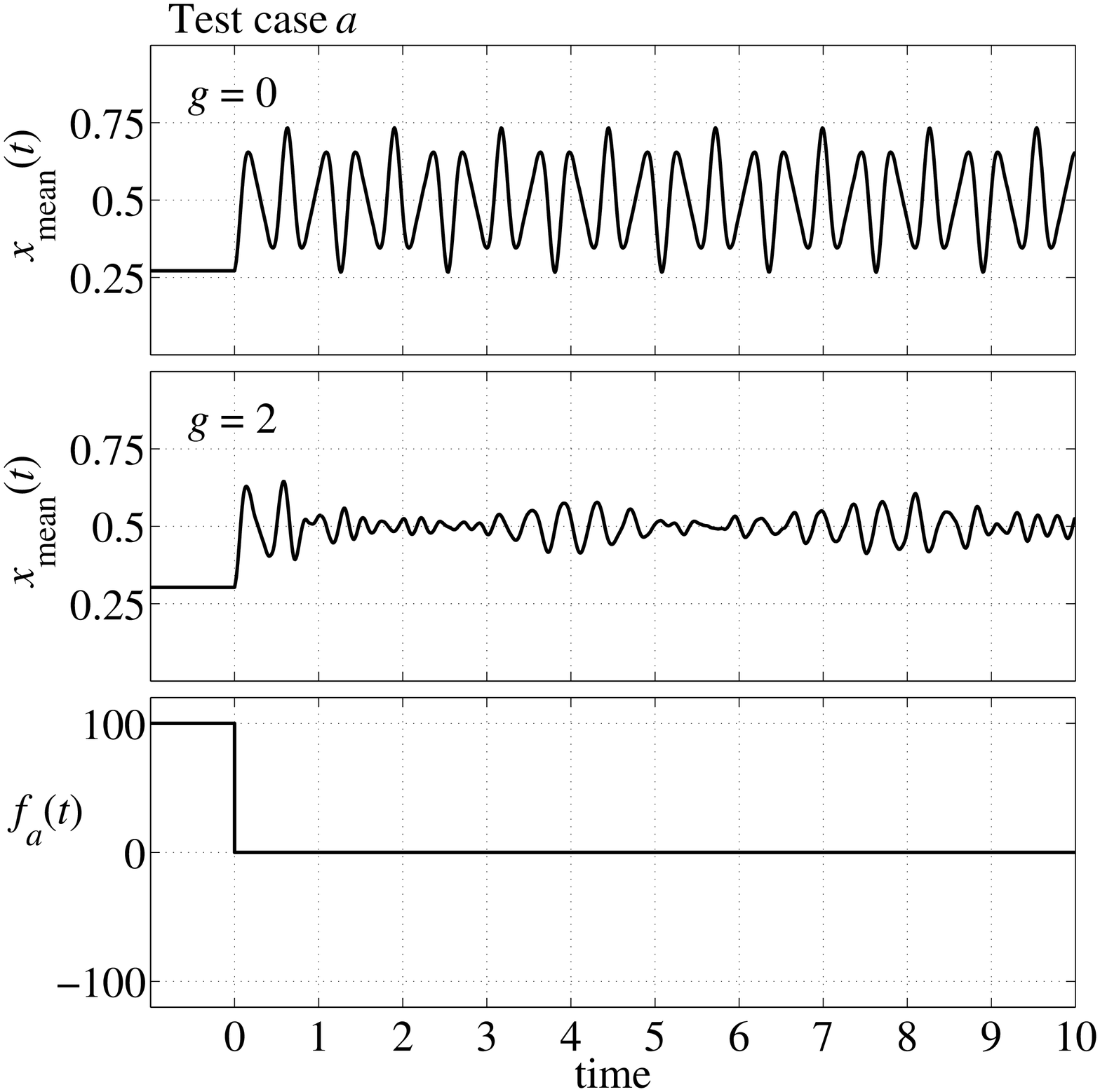}\includegraphics[width=0.5\textwidth]{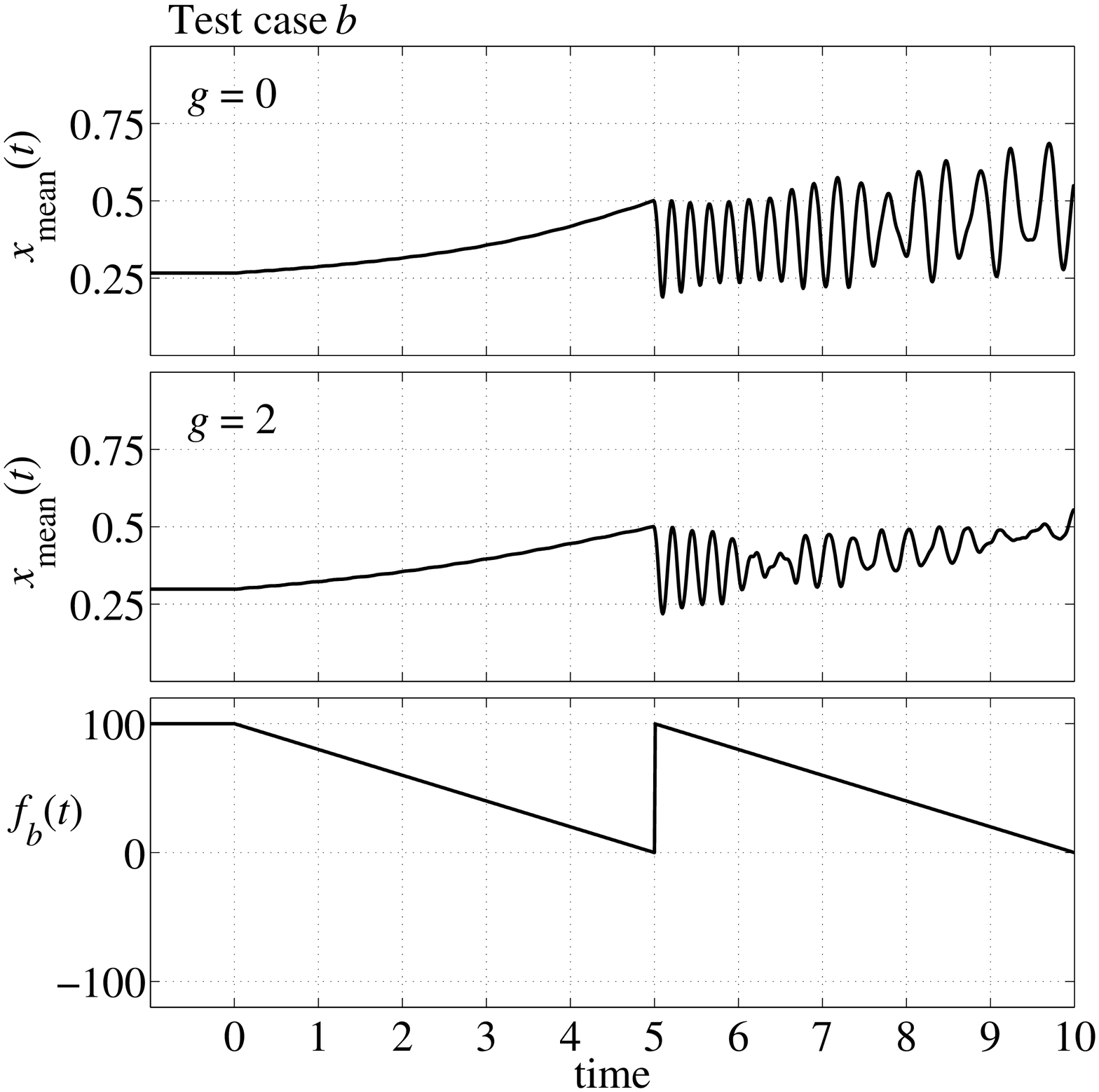}
\par\end{centering}

\noindent \begin{centering}
\includegraphics[width=0.5\textwidth]{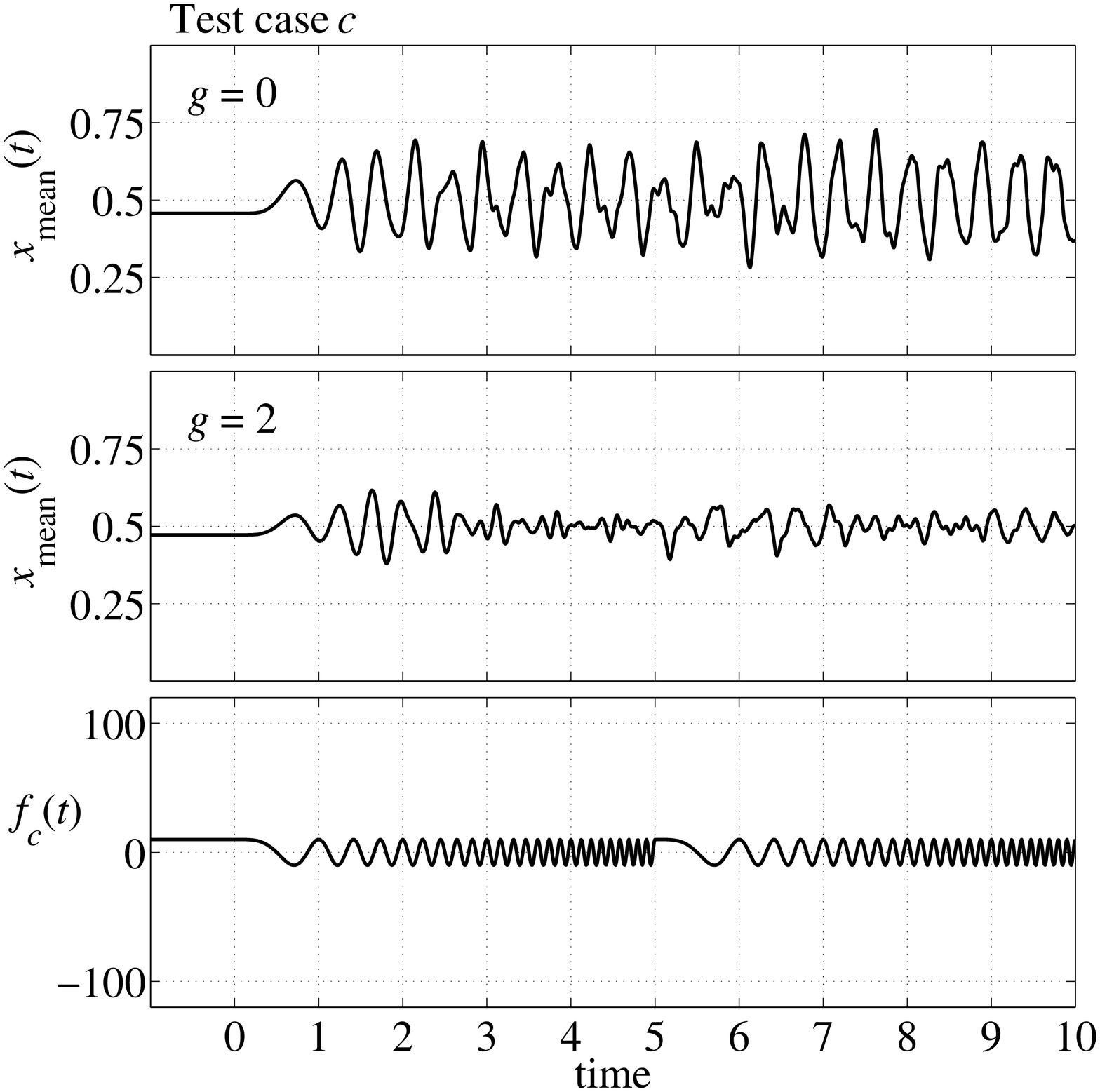}\includegraphics[width=0.5\textwidth]{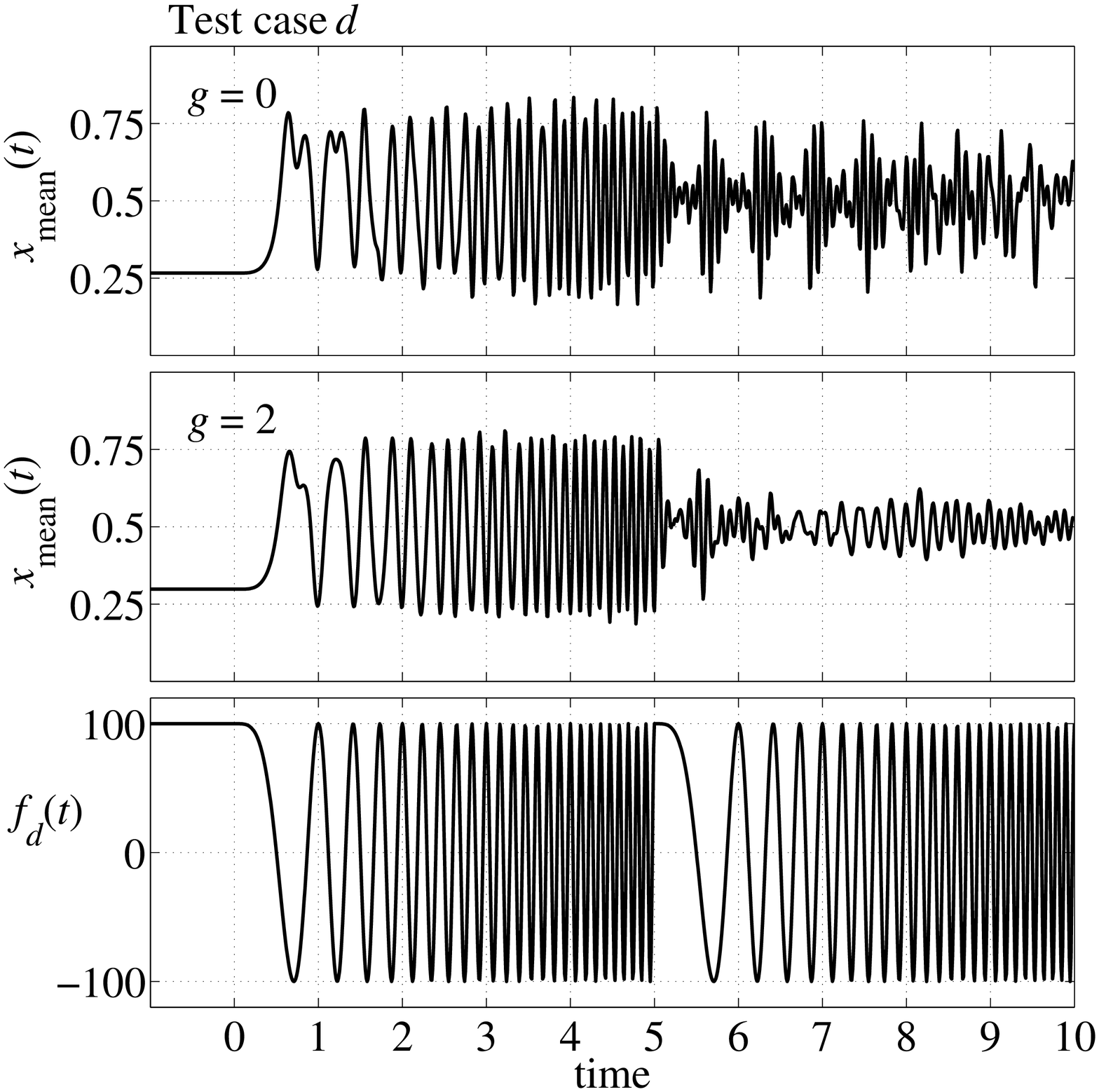}
\par\end{centering}

\caption{\label{fig:results-oscillations}\textbf{Dynamics of the test systems}
The mean position $x_{\mathrm{mean}}(t)$ of the $N=5$ particles
in the well; see details in the text. To see how the interactions
between particles affect the result, both the cases $g=0$ and $g=2$
are included. The function $f(t)$ that determines the strength of
the tilted bottom is also plotted for each case.}
\end{figure}

\par\end{center}

\subsection{\label{sub:Results-TestResults}Test results and discussion}

The various methods discussed in section \ref{sec:Methods} were tested
and compared, for the test case discussed above. To summarize, the
exponential integrator based methods (ALC, AL1 and AL2) were found
to perform well, in particular for large basis sizes, and when the
Hamiltonian itself is not changing too strongly and rapidly in time.
Details are given below.

\subsubsection{Performance measurement}

With large dimensions of the Hilbert space, most of the computational
time required is used to form matrix-vector products between the Hamiltonian
and some vector. For all the methods considered here, all other calculations
involved can be expected to take much less time, typically being scalar
products or additions of vectors. Performance is therefore here measured
in the number of such matrix-vector multiplications that a method
needs in order to produce $\vert\Psi(t=10)\rangle$.

\subsubsection{\label{sub:Results-TestResults-NumerParams}Numerical parameters}

Some remarks should be made regarding the various numerical tolerances
for the different methods. The aim in this study was to find methods
which are accurate enough, so that the global error should be smaller
than $10^{-2}$ (that is, for the present test case, $\Vert\vert\Psi_{\mathrm{approx}.}(t=10)\rangle-\vert\Psi_{\mathrm{correct}}(t=10)\rangle\Vert_{2}<10^{-2}$).
This was here assumed to be fulfilled if the solutions obtained with
different methods did not differ by more than $10^{-2}$. An error
of this order implies that expectation values of the calculated states
should be accurate enough, so that one can draw qualitative conclusions
about the underlying physics.

After extensive testing, all error tolerances used for the methods
ALC, AL1, AL2 and RK8 were eventually fixed to $10^{-6}$. The various
tolerances essentially control the local error per time step, so that
it seems reasonable that the appropriate values should be similar
for the different methods. In any case, larger values occasionally
gave too large global errors, and the aim here was to find robust
methods needing as little tuning as possible.

Regarding the parameter $d_{1}$, the number of single-particle orbitals
used in the basis, an increased $d_{1}$ gives a higher spatial accuracy.
For the present test case, the resulting $x_{\mathrm{mean}}(t)$ change
a bit when $d_{1}$ is increased from $10$ to $20$, but the difference
when increasing $d_{1}$ to $30$ is barely visible on the scale of
Fig.~\ref{fig:results-oscillations}. A stronger tilted potential,
or a stronger interaction between the particles, would require more
orbitals to be used in order to achieve accurate results. A larger
dimension $d_{1}=40$ is also included in Table \ref{tab:results-data},
but because of practical time limitations only the method AL1 was
tested for that case.

\subsubsection{\label{sub:Results-TestResults-TestResults}Test results}

Test results for the methods presented in section \ref{sec:Methods},
are given in Table \ref{tab:results-data}, for various choices of
basis sizes and time-dependent potentials. For the different cases,
the number of matrix-multiplications required by a method to produce
$\vert\Psi(t=10)\rangle$ are given in the respective columns. With
the numerical parameters set as discussed above, the error between
solutions obtained with different methods were typically of the order
$10^{-3}$.

For case (\emph{a}) with a constant Hamiltonian, the ALC method was
substantially more efficient than any of the other methods, needing
fewer matrix-vector multiplications. As one might expect from their
differences with ALC, the methods AL1 and AL2 need to do more work
to control the error -- essentially they to compute each time step
twice. In this sense there is no point in actually using those methods
for constant Hamiltonians, but the present test results imply that
the adaptivity works fairly well, given how they operate.

Generally, AL1 was more efficient than AL2. Regarding the Runge-Kutta
method (RK8), it appears to be indifferent to the time-dependence
in the Hamiltonian, such that it does about the same amount of work
for each of the four cases. Contrary to this, AL1 and AL2 have to
work harder when the time-dependence becomes more influential, as
in case \emph{d}, although they are still more efficient than RK8
for the largest basis sizes.

For all the methods, except ALC, it will occasionally happen that
a time step needs to be re-done with a smaller $\Delta t$ in order
to keep the error small enough. Thus, some matrix-vector multiplications
will have been wasted. These are included in the values reported in
Table~\ref{tab:results-data}, but typically constituted only a few
percent of the total count.

Several other choices of basis sizes, particle numbers and time-dependent
potentials were tested as well, apart from those reported in Table~\ref{tab:results-data},
and all were found to give consistent results.

\begin{table}
\begin{centering}
\begin{tabular}{ccc|cccc}
$f(t)$ & $d_{1}$ & $d_{N}$ & ALC & AL1 & AL2 & RK8\tabularnewline
\hline
 &  &  &  &  &  & \tabularnewline
\emph{a} & 10 & 2002 & 16200 & 36040 & 50622 & 87229\tabularnewline
 & 20 & 42504 & 74670 & 128850 & 182466 & 347172\tabularnewline
 & 30 & 278256 & 160050 & 297104 & 355938 & 780333\tabularnewline
 & 40 & 1086008  & -- & 568560 & -- & --\tabularnewline
 &  &  &  &  &  & \tabularnewline
\emph{b} & 10 & 2002 & -- & 64623 & 81176 & 92452\tabularnewline
 & 20 & 42504 & -- & 148960 & 165565 & 352094\tabularnewline
 & 30 & 278256 & -- & 302143 & 290137 & 785391\tabularnewline
 & 40 & 1086008  & -- & 559624 & -- & --\tabularnewline
 &  &  &  &  &  & \tabularnewline
\emph{c} & 10 & 2002 & -- & 108039 & 132063 & 87215\tabularnewline
 & 20 & 42504 & -- & 196115 & 232306 & 347160\tabularnewline
 & 30 & 278256 & -- & 320254 & 344869 & 780372\tabularnewline
 & 40 & 1086008  & -- & 558616 & -- & --\tabularnewline
 &  &  &  &  &  & \tabularnewline
\emph{d} & 10 & 2002 & -- & 301651 & 357154 & 105479\tabularnewline
 & 20 & 42504 & -- & 326627 & 410428 & 347267\tabularnewline
 & 30 & 278256 & -- & 459331 & 609458 & 780854\tabularnewline
 & 40 & 1086008  & -- & 653302 & -- & --\tabularnewline
\end{tabular}
\par\end{centering}

\caption{\label{tab:results-data}\textbf{Test results} The number of matrix-vector
multiplications required by a method to produce $\vert\Psi(t=10)\rangle$
are listed for the four different methods denoted by ALC, AL1, AL2
and RK8 (see section \ref{sec:Methods}). Tests were run for different
time-dependent potentials modulated by different functions $f(t)$
(cases \emph{a}, \emph{b}, \emph{c} and \emph{d}, see section \ref{sub:Results-TestCase-TDPotentials}),
and using different numbers $d_{1}$ of basis orbitals (see section
\ref{sub:Results-TestCase-BasisStates}). For the $N=5$ particles
in the system, the resulting Hilbert space dimension is $d_{N}$ (that
is, the matrix has dimensions $d_{N}\times d_{N}$). Since the ALC
method is only applicable for constant Hamiltonians, some cells are
empty. Also, because of practical time limitations, only the AL1 method
was tested for the case $d_{1}=40$. As discussed in section \ref{sub:Results-TestResults-NumerParams},
already $d_{1}=30$ gave sufficient spatial accuracy. The error between
solutions produced by different methods was typically of the order
$10^{-3}$, and never larger than $10^{-2}$.}
\end{table}

\subsubsection{\label{sub:Results-TestResults-ComparOthers}Comparison with other
methods}

A number of other methods were also tested. A commonly used 5:th order
Runge-Kutta method, also designed by Dormand and Prince \cite{dormand-prince-5-4},
was found to require about 70\% more matrix-vector multiplications
than the 8:th order version (see section \ref{sub:Methods-RK}), for
the present test case. In Ref.~\cite{dormand-prince-5-4} three variants
are given of the 5:th order method; in this study the one called RK5(4)7M
was found to be the most efficient of those, in agreement with what
the authors concluded.

In order to compare with existing widely used implementations, some
of the routines in the commercial software package Matlab \cite{MATLAB}
were also tested. As one should expect, that implementation of the
5:th order Dormand-Prince method (called ode45 \cite{shampine-reichelt-1997})
had almost identical performance to the one made in this study. Another
Matlab routine, called ode23 \cite{shampine-reichelt-1997}, using
a lower-order Runge-Kutta method, was found to be less efficient.
The Matlab routine ode113 \cite{shampine-reichelt-1997}, using a
multi-step Adams-Bashforth-Moulton method, was found to be almost
as efficient as the 8:th order Runge-Kutta method discussed above
(RK8). A relevant question is if any of the Matlab routines intended
for stiff problems would be more suitable, but this seems unlikely
from the results discussed in section \ref{sub:Implicit-methods},
where the implicit Crank-Nicolson method is tested.

\subsubsection{\label{sub:Results-TestResults-Discussion}Discussion}

For a constant Hamiltonian, the ALC method, originally presented by
Park and Light \cite{park-light-1986}, clearly appears to be the
best choice, from the results in Table~\ref{tab:results-data}.

When the Hamiltonian explicitly depends on time, the AL1 method was
generally found to be the most efficient. If, somehow, the two different
steps used in the AL2 method could somehow be embedded in each other,
similar to embedded Runge-Kutta formulas \cite{numerical-recipes},
then AL2 would perform better than AL1, as the numbers of multiplications
it needs could then be reduced by one third. However, despite various
schemes attempted during the course of this study, no practical way
was found.

As discussed, the time step $\Delta t$ used by the AL1 and AL2 methods
was determined adaptively. Perhaps counter-intuitively, it was of
the same order throughout the entire evolution with the sawtooth function
(case \emph{b}), even though the particles in the system have very
different dynamics in different time intervals, as shown in Fig.~\ref{fig:results-oscillations}.
In other words, the numerical work required was about the same whether
the evolution was adiabatic or not.

Another relevant issue to keep in mind is the actual development work
needed to implement the methods. From this perspective, the RK8 method
is very easy to work with, at least relative to the others. Apart
from the matrix-vector multiplications, it only requires regular vector
operations like addition and scaling. It does not rely on more sophisticated
numerical algorithms, such as eigenvalue determination. And it is
fairly straightforward to implement following the self-contained presentation
in, for example, the original Ref.~\cite{dormand-prince-8-7}.

\section{\label{sec:Summary}Summary}

In this study, various numerical methods for solving the time-dependent
Schrödinger equation have been compared, for few-body systems with
strong interactions between the particles, and with an explicitly
time-dependent Hamiltonian. Using the configuration interaction method,
one obtains a linear system of ordinary differential equations, which
in principle is a standard problem. However, the typically very large
dimensions make it important to find optimal methods.

In agreement with what has been stated by other authors (see for example
Ref. \cite{hochbruck-ostermann-2010}), so called exponential integrator
methods were found to perform better than other methods designed for
more general differential equations. For the case of a time-independent
Hamiltonian, a method described in Ref. \cite{park-light-1986} was
found to perform very well (denoted ALC above), but it is not applicable
when the Hamiltonian has an explicit time-dependence. In this case,
two extended algorithms were suggested here, which implement adaptivity
in a different way. They were found to be more efficient than all
other methods tested here.

Both the extended algorithms compute each time step twice, to obtain
two solutions where one is known to be more accurate than the other.
In short, if the difference between the solutions is small enough,
the simulation can continue, otherwise the time step is re-done with
a smaller step size. In one of the approaches the second order Magnus
approximation was used and compared with the first order variant --
this provided two solutions that could be compared (this method is
denoted AL2 above). While this scheme worked well, an alternative
implementation was found to be even more efficient; using only the
first order approximation combined with step-doubling to obtain the
more accurate solution (the approach denoted AL1 above). Regarding
other types of methods, the most efficient of those that were tested
here was an 8:th order Runge-Kutta formula, with an embedded 7:th
order formula to provide the two solutions (RK8) \cite{dormand-prince-8-7}.
While being less efficient than the exponential-integrator based methods,
it is, relative to the other methods, easier to implement. An implicit
method was also tested, as discussed in the appendix, but found to
be inefficient.

Certainly, further development and refinement may improve the performance
of the methods AL1 and AL2. One interesting point to address is if
the local error estimate can be computed in a more efficient way,
to avoid the double time evolutions, perhaps similar in fashion to
embedded Runge-Kutta formulas. Another issue is the various numerical
parameters involved; their interplay is not obvious, and more careful
tuning may improve the performance. In any case, though, the perhaps
naive choices made in this study easily resulted in methods which
performed very well.

\appendix

\section{Appendix}

\subsection{\label{sub:Implementation-details}Remarks about the implementation}

\subsubsection{Restrictions on the Hamiltonian}

While the aim here was to find methods for as general Hamiltonians
as possible, one restriction was applied. The implementation made
in this study only handles time-dependent Hamiltonians that can be
written on the form

\begin{equation}
H(t)=A+f(t)B\label{eq:Ha-fHb}\end{equation}
such that the matrices $A$ and $B$ are constant throughout the time
evolution, and the entire time dependence is contained in the scalar
function $f(t)$. This form still allows one to simulate a number
of relevant physical systems, such as electrons in a laser field,
or ultra-cold atoms with time-dependent interactions governed by an
external field. No assumption is made about the relative magnitudes
of the two matrices, or whether they are for example one- or two-body
operators. In principle, both matrices could have time-dependent prefactors,
and there could be additional similar terms. The important thing is
to be able to form the product between the matrix $\Omega_{1}$ or
$\Omega_{2}$ with an arbitrary vector (see Eqs.~(\ref{eq:Omega1})
and (\ref{eq:Omega2})), at any time $t$, and for any $\Delta t$.
With the restriction in Eq.~(\ref{eq:Ha-fHb}) above, and with Simpson's
rule to approximate the integrals, these matrices reduce to

\begin{equation}
\Omega_{1}(t,\Delta t)\approx-i\Delta tA-i\frac{\Delta t}{6}\biggl(f(t)+4f(t+\frac{\Delta t}{2})+f(t+\Delta t)\biggl)B\label{eq:Omega1-reduced}\end{equation}
and

\begin{equation}
\Omega_{2}(t,\Delta t)\approx\frac{(\Delta t)^{2}}{12}\biggl(f(t+\Delta t)-f(t)\biggl)[A,B]\label{eq:Omega2-reduced}\end{equation}
so that one only needs to form the multiplication of $A$ or $B$
with arbitrary vectors, and then scale the products.

\subsubsection{Breakdown of the Lanczos process}

In the Lanczos process, when a Krylov space and the associated tridiagonal
matrix is generated -- see Eq.~(\ref{eq:lanczos-factorization})
-- it may happen that some $\beta$-value is zero. This implies that
a subspace has been found from which the state $\vert\Psi(t)\rangle$
cannot escape, which is fortunate since it makes the time evolution
easier. A typical example would be if the Hamiltonian does not change
in time, and the initial state happens to be an eigenstate. While
such a breakdown does not cause any fundamental problem for the time
evolution, it needs to be handled in an implementation.

\subsubsection{Initial guess for $\Delta t$}

Given some time interval $[t_{\mathrm{begin}},t_{\mathrm{end}}]$
during which the time evolution is to be calculated, some initial
guess for the time step $\Delta t$ is needed in the methods. In the
present implementation this was made by simply starting with $\Delta t=t_{\mathrm{end}}-t_{\mathrm{begin}}$,
and then let the methods adjust it themselves, as described in section
\ref{sec:Methods}.

\subsection{\label{sub:Implicit-methods}An implicit method}

Typically, the eigenvalues of the Hamiltonian matrix may be very different
in magnitude, because of the different energies of the basis states
used. This implies that the problem may be stiff, so that an implicit
solver might be more efficient than an explicit one \cite{numerical-recipes}.
In an initial phase of this study, the Crank-Nicolson (CN) method
\cite{crank-nicholson-1947} was tested. It is an implicit method
similar to the Euler backward method, such that it is stable, but
of higher accuracy. Given the state $\vert\Psi(t_{0})\rangle$, the
next state $\vert\Psi(t_{0}+\Delta t)\rangle$ is defined implicitely
by the equation

\begin{equation}
\biggl(1+i\frac{\Delta t}{2}H(t_{0}+\Delta t)\biggl)\vert\Psi(t_{0}+\Delta t)\rangle=\biggl(1-i\frac{\Delta t}{2}H(t_{0})\biggl)\vert\Psi(t_{0})\rangle\label{eq:crank-nicholson}\end{equation}
To obtain the state $\vert\Psi(t_{0}+\Delta t)\rangle$, one here
needs to solve a system of linear equations. For this purpose, the
generalized minimum residual method (GMRES) \cite{golub-vanloan,saad-schultz-1986,templates-for-linear-systems}
was used here. It is an iterative method which, given an initial guess,
converges towards the solution by use of repeated matrix-vector multiplications,
making it suitable to use with large and sparse matrices. For the
present application the method typically converged within 5--10 iterations
(here, this was when the residual was less than $10^{-6}$). The Euler
forward method was used to generate the initial guess.

However, it appears that the physical problem studied here was not
sufficiently stiff to make an implicit method competitive. Or, perhaps,
that so high accuracy is required that a higher-order method is needed
(as also discussed in Ref.~\cite{hochbruck-ostermann-2010} for this
class of problems). Table \ref{tab:results-RK-vs-CN} shows some results
from a comparison with the (explicit) classical fourth-order Runge-Kutta
method \cite{numerical-recipes}, for the test case above with $N=2$
particles. As discussed in section \ref{sub:Results-TestResults},
performance is here measured in the number of matrix-vector multiplications
needed. For this test, the methods were not implemented to be adaptive,
and instead a manually chosen fixed time step $\Delta t$ was used
throughout the simulation. The obtained solution was then compared
with that produced by the adaptive Lanczos solver described above
for constant Hamiltonians (ALC). The time step was adjusted to give
an error less than $0.01$, and to limit the available parameter space
only time steps defined by two significant digits were considered
(for example $\Delta t=0.014$ or $0.015$, but not $0.0145$). As
shown in Table \ref{tab:results-RK-vs-CN}, the Crank-Nicolson method
required considerably more multiplications than the explicit method.

\begin{table}
\begin{centering}
\begin{tabular}{cc|cc|cc}
 &  & RK4 &  & CN & \tabularnewline
$d_{1}$ & $d_{N}$ & mults. & $\Delta t_{\mathrm{RK4}}$ & mults. & $\Delta t_{\mathrm{CN}}$\tabularnewline
\hline
20 & 210 & 56340 & 0.00071 & 285797 & 0.00014\tabularnewline
40 & 820 & 235296 & 0.00017 & 800003 & 0.000050\tabularnewline
\end{tabular}
\par\end{centering}

\caption{\label{tab:results-RK-vs-CN}Test results for the explicit fourth-order
Runge Kutta method (RK4) and the implicit Crank-Nicolson (CN); see
details in the text, and see also the explanation of captions in Table
\ref{tab:results-data}. The constant Hamiltonian from test case \emph{a}
was considered here, with $N=2$ particles. The CN method requires
considerably more multiplications than the RK4 method. When comparing
with the results in Table \ref{tab:results-data}, it should be noted
that the numbers here do not include any cost for adaptivity, as the
fixed time step $\Delta t$ was adjusted manually. The error is here
defined as $\Vert\Psi_{\mathrm{RK4/CN}}-\Psi_{\mathrm{ALC}}\Vert_{2}$,
the difference with the solution obtained using the ALC method. Several
other choices of basis sizes, particle numbers and time-dependent
potentials were tested, and all were found to give very similar results,
also for much larger problem dimensions.}
\end{table}

\subsubsection*{Acknowledgements}

This work was supported by the Swedish Research Council and the Nanometer
Structure Consortium at Lund University.

\end{document}